\pdfoutput=1
\documentclass[journal]{IEEEtran}
\usepackage{cite}
\usepackage[utf8]{inputenc}
\usepackage{graphicx}
\usepackage{amssymb}
\usepackage[usenames, dvipsnames]{xcolor}
\usepackage{footmisc}
\usepackage{booktabs}
\usepackage{multirow}
\usepackage{pgfplots}
\pgfplotsset{compat=newest}
\usepackage{tikz}
\usepackage{url}
\usepackage{xspace}
\usepackage{tabularx}
\usepackage{hyperref}
\usepackage{enumitem}
\usepackage{multicol}
\usepackage{array}
\usepackage{balance}
\usepackage{orcidlink}

\usepackage[reqno]{amsmath}
\makeatletter
\newcommand{\leqnomode}{\tagsleft@true\let\veqno\@@leqno}
\newcommand{\reqnomode}{\tagsleft@false\let\veqno\@@eqno}
\makeatother

\usepackage{nccmath}
\usepackage{mathtools}

\newcolumntype{L}[1]{>{\raggedright\let\newline\\\arraybackslash\hspace{0pt}}m{#1}}
\newcolumntype{C}[1]{>{\centering\let\newline\\\arraybackslash\hspace{0pt}}m{#1}}
\newcolumntype{R}[1]{>{\raggedleft\let\newline\\\arraybackslash\hspace{0pt}}m{#1}}

\usetikzlibrary{dsp, chains, decorations.markings, positioning, shapes.geometric, shadows, fit,
backgrounds}

\pgfplotsset{
	legend entry/.initial=,
	every axis plot post/.code={%
		\pgfkeysgetvalue{/pgfplots/legend entry}\tempValue
		\ifx\tempValue\empty
		\pgfkeysalso{/pgfplots/forget plot}%
		\else
		\expandafter\addlegendentry\expandafter{\tempValue}%
		\fi
	},
}

\usepackage{subcaption}

\definecolor{mcma}{RGB}{125, 125, 255}
\definecolor{mcmafill}{RGB}{200, 200, 255}
\definecolor{mcmad}{RGB}{255, 100, 100}
\definecolor{mcmadfill}{RGB}{255, 130, 130}
\definecolor{mcmb}{RGB}{255, 127, 0}
\definecolor{mcmbfill}{RGB}{255, 167, 50}
\definecolor{tmcm}{RGB}{0, 150, 0}
\definecolor{tmcmfill}{RGB}{0, 190, 0}
\definecolor{mcmadmax}{RGB}{255, 100, 100}
\definecolor{mcmadfillmax}{RGB}{255, 160, 160}
\definecolor{mcmbmax}{RGB}{255, 127, 0}
\definecolor{mcmbfillmax}{RGB}{255, 227, 100}
\definecolor{tmcmmax}{RGB}{0, 150, 0}
\definecolor{tmcmfillmax}{RGB}{0, 230, 0}

\newcommand{\intent}[2]{[\![#1; #2]\!]}

\newcommand{\bool}{\left\{0, 1\right\}}
\newcommand{\sh}{\text{\normalfont sh}}
\newcommand{\nsh}{\text{\normalfont nsh}}
\newcommand{\sg}{\text{\normalfont sg}}
\newcommand{\odd}{\text{\normalfont odd}}

\newcommand{\AD}{\text{\normalfont AD}}

\newcommand{\Smin}{S_{\min}}
\newcommand{\Smax}{S_{\max}}
\newcommand{\msb}{\text{\normalfont msb}}
\newcommand{\OB}{\text{\normalfont B}}

\newcommand{\eg}{\emph{e.\,g.}\xspace}
\newcommand{\ie}{\emph{i.\,e.}\xspace}

\newcommand{\wrt}{{w.\,r.\,t.}\xspace}

\newcommand{\ad}{\mathit{ad}}
\newcommand{\bNA}{\overline{N_A}}
\newcommand{\NA}{N_A}
\newcommand{\MCMmAD}{MCM$_{\min\AD}$\xspace}
\newcommand{\MCM}{MCM\xspace}
\newcommand{\MCMA}{MCM-Adders\xspace}
\newcommand{\MCMB}{MCM-Bits\xspace}
\newcommand{\TMCM}{tMCM\xspace}
\newcommand{\MCMad}{MCM$_{\AD}$\xspace}

\newcommand{\lr}{\left\{l,r\right\}}

\newcommand{\gainLUTsOBAvsNA}{7.59}

\newcommand{\gainOBAsTvsOBA}{25.31}

\newcommand{\gainLUTsTvsOBA}{21.55}
\newcommand{\gaindelayTvsOBA}{5.15}
\newcommand{\gainpowerTvsOBA}{18.68}
\newcommand{\gainOBAsThalfvsOBA}{28.87}
\newcommand{\gainOBAsTquartervsOBA}{6.2}

\newcommand{\gainLUTsThalfvsOBA}{28.87}
\newcommand{\gainLUTsThalfvsNA}{35.93}

\usepackage{soul}

\usepackage{todonotes}

\begin{document}
\reqnomode

\title{Towards the Multiple Constant Multiplication at Minimal Hardware Cost}
\author{R\'{e}mi Garcia\orcidlink{0000-0001-6704-759X} and Anastasia Volkova\orcidlink{0000-0002-0702-5652}%
\thanks{R. Garcia and A. Volkova were with Nantes Université, CNRS, LS2N, 44000 Nantes. Email:~firstname.lastname@univ-nantes.fr}%
}

\maketitle

\begin{abstract}
Multiple Constant Multiplication (MCM) over integers is a frequent operation arising in embedded systems that require highly optimized hardware. An efficient way is to replace costly generic multiplication by bit-shifts and additions, \ie{} a multiplierless circuit.
In this work, we improve the state-of-the-art optimal approach for MCM, based on Integer Linear Programming (ILP). We introduce a new lower-level hardware cost, based on counting the number of one-bit adders and demonstrate that it is strongly correlated with the LUT count.
This new model for the multiplierless MCM circuits permitted us to consider intermediate truncations that permit to significantly save resources when a full output precision is not required. We incorporate the error propagation rules into our ILP model to guarantee a user-given error bound on the MCM results.
The proposed ILP models for multiple flavors of MCM are implemented as an open-source tool and, combined with the FloPoCo code generator, provide a complete coefficient-to-VHDL flow.
We evaluate our models in extensive experiments, and propose an in-depth analysis of the impact that design metrics have on actually synthesized hardware.
\end{abstract}


\IEEEpeerreviewmaketitle

\section{Introduction}

Multiplications by integer constants arise in many numerical algorithms and applications.
In particular, algorithms that target embedded systems often involve Fixed-Point (FxP) numbers which can be assimilated to integers.
These algorithms range from dot-product evaluation for deep neural networks to more complex algorithms such as digital filters.

In order to save hardware resources, the knowledge on the constants to multiply with can be used in dedicated multiplierless architectures, instead of costly generic multipliers~\cite{WiatrJamro_Constantcoefficientmultiplication_2000}.
The shift-and-add approach is the privileged method to reduce hardware cost, it consists in replacing multiplications by additions/subtractions and bit-shifts, which are multiplications of the data by a power of two that can be hardwired for a negligible cost.
For example, multiplying an integer variable $x$ by the constant $7$ can be rewritten as $7x = 2^3x - x$, reducing the cost to a single bit-shift by three positions to the left and a subtraction, instead of a multiplication.

\begin{figure}
	\centering
	\begin{subfigure}{0.37\linewidth}
		\centering
		\includegraphics[width=1.0\linewidth]{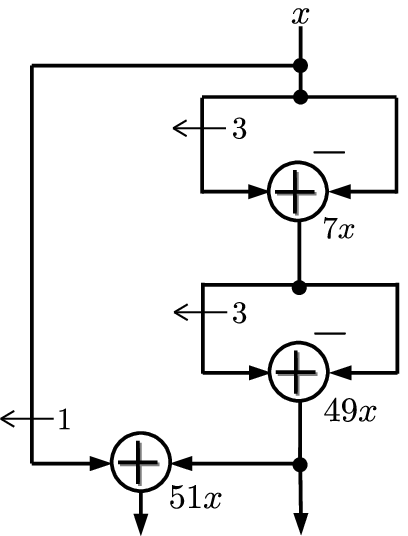}
		\caption{$AD = 3$}\label{fig:4951high}
	\end{subfigure}
	\hspace{0.08\linewidth}
	\begin{subfigure}{0.37\linewidth}
		\centering
		\includegraphics[width=1.0\linewidth]{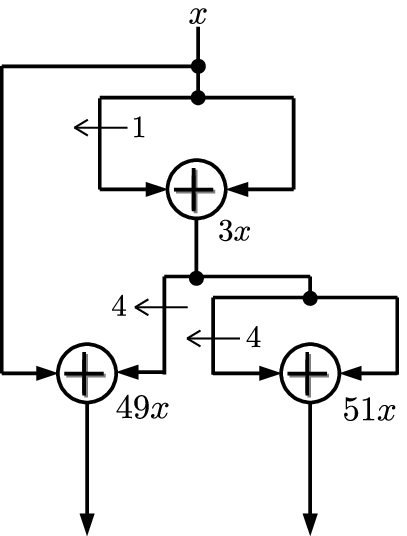}
		\caption{$AD = 2$}\label{fig:4951low}
	\end{subfigure}
	\caption{Different adder graph topologies computing the same outputs: $49x$ and $51x$}\label{fig:compareonebit}
\end{figure}

Given a set of target constants to multiply with, finding the implementation with the lowest cost is called the Multiple Constant Multiplication (\MCM) problem.
Typical way to tackle the problem is to find shift-and-add implementations represented using \emph{adder graphs}, as in \figurename{}~\ref{fig:compareonebit}, that describe the multiplierless solutions with the \emph{minimum number of adders}. In the following, this problem will be referred to as the \MCMA. The main objective of this work is to first improve the existing approaches for the \MCMA problem, then push towards finer-grained hardware cost metrics counting number of \textit{one-bit adders} (the \MCMB problem), and finally, use truncations in internal data paths to considerably save resources (the \TMCM problem).

It is straightforward to obtain a first shift-and-add solution from the binary representation of the constant.
A greedy algorithm based on the Canonical Signed Digit (CSD) representation permits to reduce the number of adders \cite{Booth_SignedBinaryMultiplication_1951, Bernstein_Multiplicationintegerconstants_1986}.
Many heuristics enhance the results obtained with the CSD method~\cite{DempsterMacleod_Constantintegermultiplication_1994, PotkonjakSrivastavaChandrakasan_Multipleconstantmultiplications_1996, Lefevre_MultiplicationIntegerConstant_2001, VoronenkoPueschel_Multiplierlessmultipleconstant_2007, ThongNicolici_Combinedoptimalheuristic_2010}, but heuristics do not provide any guarantee on the solution quality.
Optimal approaches for the \MCMA can be roughly divided into two categories: (i) first approaches based on hypergraphs \cite{Gustafsson_Towardsoptimalmultiple_2008} and Integer Linear Programming (ILP) model~\cite{Kumm_MultipleConstantMultiplication_2016_book}, which can be solved by generic solvers; (ii) dedicated optimal algorithms based on branch and bound technique proposed and further developed by Aksoy et al.~\cite{AksoyGuenesFlores_Searchalgorithmsmultiple_2010}. In this work, we improve and extend the results of the first category of approaches, since ILP-based modeling offers a better versatility for extensions than dedicated algorithms, and permits to rely on the efficiency of generic solvers.

The state-of-the-art ILP-based model for the \MCMA was proposed by Kumm in~\cite{Kumm_OptimalConstantMultiplication_2018}.
This method finds optimal solutions in terms of the number of adders in reasonable time, and has been adapted to SAT/SMT solvers \cite{DerivingOptimalMultiplication_2020} for single constant multiplication.
However, we found an error in the model which makes it miss some optimal solutions.

In this work, we use~\cite{Kumm_OptimalConstantMultiplication_2018} as basis, correct the modeling error for the \MCMA problem and build upon it to solve other flavors of the MCM.
In particular, the number of cascaded adders, called the \emph{adder depth} (AD), directly impacts the delay of the circuit, and is hence desired to be bounded.
We first encode the AD count in our ILP model, and secondly propose, for the first time, a new bi-objective formulation called \MCMad, which minimizes the number of adders and the AD \textit{simultaneously}.

One of the main contributions of this paper is solving the MCM for a low-level hardware metric based on counting the one-bit adders, when the word length of the input $x$ is known \emph{a priori}.
Assume adder graphs in \figurename~\ref{fig:compareonebit} taking a 3-bit input $x$. While both require only three adders, the solution in \figurename~\ref{fig:4951high} requires 22 one-bit adders and the other in \figurename~\ref{fig:4951low} uses only 9.
While classical bit-level optimization approaches are iterative and require synthesis and simulations \cite{DempsterMacleod_Constantintegermultiplication_1994, JohanssonGustafssonWanhammar_BitLevelOptimization_2007}, solving the MCM with a fine-grained cost model is done only once.
By one-bit adders, we gather both half and full adders as logic elements having roughly the same cost.
This low-level metric has been discussed for decades, see \cite{DempsterMacleod_Constantintegermultiplication_1994} in which a heuristic for fixing the topology was presented. To our knowledge, we propose the first optimal approach solving the \MCM problem targeting the one-bit adders (the \MCMB problem).

We further extend \MCMB model to solve the Truncated MCM (\TMCM) problem.
Indeed, the output results of the computations do not necessarily need to be full-precision and in practice are often rounded post-MCM.
Focusing on a low-level metric allows to add intermediate truncations \cite{GuoDeBrunnerJohansson_TruncatedMCMusing_2010, DinechinFilipKummForget_TableBasedversus_2019, GarciaVolkovaKumm_TruncatedMultipleConstant_2022} that will save resources and not waste area and time to compute bits that will have no impact on the rounded result. At the same time, our goal is to respect a user-given absolute error on the result.

Some previous works address the \TMCM but applied to a \emph{fixed} adder-graph and either do not give a guarantee on the output error, \eg{}, the heuristic-based approach \cite{GuoDeBrunnerJohansson_TruncatedMCMusing_2010}, or overestimate and sometimes wrongly-compute the output error, \eg{}, ILP-based \cite{DinechinFilipKummForget_TableBasedversus_2019}. However, as demonstrated in \figurename~\ref{fig:4951trunc}, some adder graph topologies are better suited for truncations than others. Hence, solving directly for \TMCM and delegating the design exploration to an ILP solver, is a better approach.
In this paper, we extend our preliminary work \cite{GarciaVolkovaKumm_TruncatedMultipleConstant_2022} for combining, for the first time, adder graph optimization with internal truncations. In particular, we tighten the error bounds and treat several corner cases, compared to \cite{GarciaVolkovaKumm_TruncatedMultipleConstant_2022}.

With this paper, we aim at providing a tool that democratizes access to \MCM at minimal hardware cost.
We mix the optimization techniques to model the problem, have an in-depth look at the hardware addition to provide fine-grain cost metrics and provide a sound error-analysis to give numerical guarantees on the computed output.
The main ideas of our approach are presented in Section~\ref{sec:birdview} and in the next sections we go through the details of the ILP models.
In Section~\ref{sec:expe}, we demonstrate the efficiency of our approach providing optimization results and obtained hardware comparisons.

\section{Bird view}\label{sec:birdview}

Our approach is based on ILP modeling, which consists in stating objectives and constraints as \emph{linear} equations involving integer and binary variables.
We chose to limit our possibilities to linear equations because it allows using efficient and robust solving approaches embedded in commercial or open-source solvers such as CPLEX, Gurobi, GLPK, etc.
Our end goal is to provide a tool which, given the target constants to multiply with, the choice of a cost function and several associated parameters, builds the corresponding ILP model. Solving the model with your favorite solver results in an adder graph description, which can be passed to the fixed- and floating-point core generator FloPoCo~\cite{DinechinPasca_DesigningCustomArithmetic_2011} to generate the VHDL code implementing the MCM circuit.

From the modeling point of view, we search for an adder graph that computes the product of the input $x$ by given target constants.
In this work, we center the model of an adder graph around the adders and their associated integer values, called \emph{fundamentals}.
The rules of construction of an adder graph are simple:
\begin{itemize}
	\item for every target constant, there exists one fundamental equal to its value;
	\item every fundamental is a sum of its signed and potentially shifted left and right inputs, which can be other fundamentals or the input $x$;
	\item every used fundamental can be traced back to the input $x$, meaning that the adder graph topology holds.
\end{itemize}

The search-space for the fundamentals is deduced from the target constants: a tight upper bound on the number of fundamentals can be deduced using heuristics~\cite{VoronenkoPueschel_Multiplierlessmultipleconstant_2007, KummZipfFaustChang_Pipelinedaddergraph_2012}; and the maximum value fundamentals can take is often restricted in practice by the word length of the largest coefficient.
To solve the \MCMA problem correctly, we encode the above rules as linear constraints.
As a result of resolution, we obtain a sequence of fundamentals, \eg{}, for \figurename~\ref{fig:4951high} this sequence is $\left\{1, 7, 49, 51\right\}$.

On top of that, for the \MCMB and \TMCM problems, each adder has an associated one-bit adder cost which must be correctly computed depending on the values of the inputs. Furthermore, for the \TMCM problem, for each adder we associate potential truncations for its left and right operands which permit to reduce the adder cost.
However, as each truncation induces an error, the model of error propagation should be incorporated as a set of constraints and the output errors is bounded by a user-given parameter. The challenge is to consider all possible cases and not overestimate the one-bit adders cost and errors.

The main challenges are to translate these objective functions and constraints into \emph{linear} equations over integer/binary variables.
In the following, we give details for that process.

\section{Optimal Multiple Constant Multiplication Counting Adders}\label{sec:mcm}

\subsection{The base model for \MCM}\label{sec:mcmmodel}

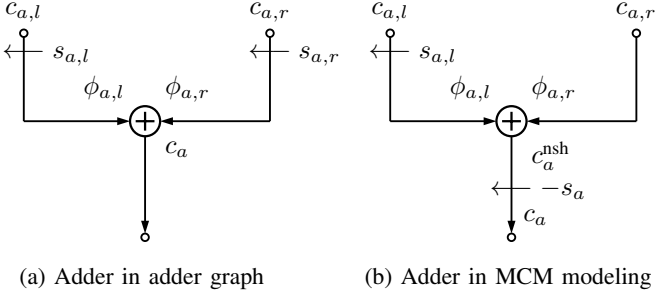
\begin{figure}
	\centering
	\begin{subfigure}{0.45\linewidth}
		\centering
		\begin{tikzpicture}\hspace{-0.7cm}
\matrix (adderclassic) [row sep=-1mm, column sep=-5mm]
{
	\node[dspnodeopen,dsp/label=above, label={below:$\phantom{s_{a,l}}\longleftarrow s_{a,l}$}] (m11) {$c_{a,l}$}; \pgfmatrixnextcell
	\node[coordinate] (m12) {}; \pgfmatrixnextcell
	\node[dspnodeopen,dsp/label=above, label={below:$\phantom{s_{a,r}}\longleftarrow s_{a,r}$}] (m13) {$c_{a,r}$}; \\
	\node[coordinate] (m21) {}; \pgfmatrixnextcell
	\node[dspadder, label={above right:$\phi_{a,r}$}, label={above left:$\phi_{a,l}$}, label={below right:$c_a$}] (m22) {}; \pgfmatrixnextcell
	\node[coordinate] (m23) {}; \\
	\node[coordinate] (m31) {}; \pgfmatrixnextcell
	\node[coordinate, label={$\phantom{-s_{a}\longleftarrow -s_{a}}$}] (m32) {}; \pgfmatrixnextcell
	\node[coordinate] (m33) {}; \\
	\node[coordinate] (m41) {}; \pgfmatrixnextcell
	\node[dspnodeopen, label={above right:$\phantom{c^{\nsh}_a}$}] (m42) {}; \pgfmatrixnextcell
	\node[coordinate] (m43) {}; \\
};
\begin{scope}[start chain]
\chainin (m11);
\chainin (m21) [join=by dspline];
\chainin (m22) [join=by dspconn];
\end{scope}
\begin{scope}[start chain]
\chainin (m13);
\chainin (m23) [join=by dspline];
\chainin (m22) [join=by dspconn];
\end{scope}
\begin{scope}[start chain]
\chainin (m22);
\chainin (m42) [join=by dspconn];
\end{scope}
\end{tikzpicture}

 
		\caption{Adder in adder graph}\label{fig:mcmadder}
	\end{subfigure}\hfill
	\begin{subfigure}{0.45\linewidth}
		\centering
		\begin{tikzpicture}\hspace{-0.7cm}
	\matrix (adderclassic) [row sep=-1mm, column sep=-5mm]
	{
		\node[dspnodeopen,dsp/label=above, label={below:$\phantom{s_{a,l}}\longleftarrow s_{a,l}$}] (m11) {$c_{a,l}$}; \pgfmatrixnextcell
		\node[coordinate] (m12) {}; \pgfmatrixnextcell
		\node[dspnodeopen,dsp/label=above, label={below:$\phantom{s_{a,r}\longleftarrow s_{a,r}}$}] (m13) {$c_{a,r}$}; \\
		\node[coordinate] (m21) {}; \pgfmatrixnextcell
		\node[dspadder, label={above right:$\phi_{a,r}$}, label={above left:$\phi_{a,l}$}, label={below right:$c^{\nsh}_a$}] (m22) {}; \pgfmatrixnextcell
		\node[coordinate] (m23) {}; \\
		\node[coordinate] (m31) {}; \pgfmatrixnextcell
		\node[coordinate, label={$\phantom{-s_{a}}\longleftarrow -s_{a}$}] (m32) {}; \pgfmatrixnextcell
		\node[coordinate] (m33) {}; \\
		\node[coordinate] (m41) {}; \pgfmatrixnextcell
		\node[dspnodeopen, label={above right:$c_a$}] (m42) {}; \pgfmatrixnextcell
		\node[coordinate] (m43) {}; \\
	};
	\begin{scope}[start chain]
		\chainin (m11);
		\chainin (m21) [join=by dspline];
		\chainin (m22) [join=by dspconn];
	\end{scope}
	\begin{scope}[start chain]
		\chainin (m13);
		\chainin (m23) [join=by dspline];
		\chainin (m22) [join=by dspconn];
	\end{scope}
	\begin{scope}[start chain]
		\chainin (m22);
		\chainin (m42) [join=by dspconn];
	\end{scope}
\end{tikzpicture}
		\vspace{-1.2em}
		\caption{Adder in MCM modeling}\label{fig:newmcmadder}
	\end{subfigure}
	\caption{Classic (left) and proposed (right) adder models}\label{fig:addermodels}
\end{figure}

The main challenge behind this model is to be able to formally state the constraints fixing the adder graph topology with linear equations.
For conciseness, we also extensively use the so-called indicator constraints, which are supported by most modern MILP solvers, and have the form of:
\begin{equation}\label{eq:indicatorcst}
	a x \leq b \quad \text{ if } y = 1,
\end{equation}
where $a$ and $b$ are constants, $x$ is an integer variable and $y$ is a binary variable.
These indicator constraints can be used as is, or linearized using big M constraints which transform \eqref{eq:indicatorcst} into
\begin{equation}\label{eq:bigmcst}
	a x \leq b + M \times \left(1-y\right),
\end{equation}
where $M$ is a constant large enough such that if $y=0$ then \eqref{eq:bigmcst} is similar to $ax\leq \infty$.
Note that indicator constraints, when passed as is, to common MILP solvers, might slow the solving process.
On the other hand, big M constraints are typically faster but could lead to numerical instabilities and should be used carefully when the value of $M$ is order of magnitudes higher than $ax$ \cite{KlotzNewman_Practicalguidelinessolving_2013, BelottiBonamiFischettiLodiMonaciNogalesGomezSalvagnin_handlingindicatorconstraints_2016}.

To fix the adder-graph topology with linear constraints, we first define a few sets of variables encoding adder information.
Integer variables $c_a$ correspond to the fundamentals for each adder and $c_{a,i}$, $\forall i \in \lr$, are the left and right inputs of the adder $a$.
To simplify, the adder graph input is handled as the zeroth adder and its associated fundamental is one: $c_0 = 1$.
Variables encoding shifts and signs are $s_{a,i}$ and $\phi_{a,i}$, respectively.
\figurename~\ref{fig:mcmadder} represents an adder and these variables which are linked together through the relation
\begin{equation}\label{eq:adderrel}
c_a = \left(-1\right)^{\phi_{a,l}}2^{s_{a,l}}c_{a,l}+\left(-1\right)^{\phi_{a,r}}2^{s_{a,r}}c_{a,r}.
\end{equation}
It has been proven \cite{DempsterMacleod_Constantintegermultiplication_1994} that we can limit the adder graph to odd fundamentals only, limiting the relevant shifts.
Thus, it is possible to simplify the modeling by considering only positive left shifts, as illustrated by \figurename~\ref{fig:newmcmadder}, and let
\begin{equation}\label{eq:newadderrel}
c_a = 2^{-s_a}\left(\left(-1\right)^{\phi_{a,l}}2^{s_{a,l}}c_{a,l}+\left(-1\right)^{\phi_{a,r}}c_{a,r}\right),
\end{equation}
where the shifts, $s_a$ and $s_{a,l}$, take values ensuring that $c_a$ is odd.
Previous work \cite{Kumm_OptimalConstantMultiplication_2018}, did consider positive left shift only or identical negative shifts, which is equivalent to our $s_a$.
However, in their linearized model, integrity constraints conflicted with negative shifts dropping the latter when solving.
Nevertheless, negative shifts are absolutely necessary to find optimal solution in some cases, \eg{}, for the target constants $C=\left\{7, 19, 31\right\}$ where $19 = 2^{-1}\left(7+31\right)$.

\begin{table}
	\centering
	\begin{tabularx}{\linewidth}{X}
		\toprule
		Constants/Variables and their meaning \\
		\midrule
		$\bNA \in \mathbb{N}$: bound on the number of adders; \\
		$N_O \in \mathbb{N}$: number of outputs; \\
		$C \in \mathbb{N}^{N_O}$: odd target constants; \\
		$w \in \mathbb{N}$: fundamentals' word length; \\
		$\Smax \in \mathbb{N}$: maximum shift, $\Smax = w$ by default. \\
		\midrule
		$c_a \in \intent{0}{2^w}$, $\forall a \in \intent{0}{\bNA}$: fundamental, or constant, obtained in adder $a$ with $c_0$ fixed to the value $1$, corresponding to the input; \\
		$c_a^\nsh \in \intent{0}{2^{w+1}}$, $\forall a \in \intent{1}{\bNA}$: constant obtained in adder $a$ before the negative shift; \\
		$c_a^\odd \in \mathbb{N}$, $\forall a \in \intent{1}{\bNA}$: variable used to ensure that $c_a$ is odd; \\
		$c_{a,i} \in \intent{0}{2^w}$, $\forall a \in \intent{1}{\bNA}$, $i \in \left\{l, r\right\}$: constant of adder from input $i$ before adder $a$; \\
		$c_{a,l}^{\sh} \in \intent{0}{2^{w+1}}$, $\forall a \in \intent{1}{\bNA}$: constant of adder from left input before adder $a$ and after the left shift; for simplification $c_{a,r}^{\sh}$ is an alias of $c_{a,r}$; \\
		$c_{a,i}^{\sh, \sg} \in \intent{-2^{w+1}}{2^{w+1}}$, $\forall a \in \intent{1}{\bNA}$, $i \in \left\{l, r\right\}$: signed constant of adder from input $i$ before adder $a$ and after the shift; \\
		$\Phi_{a,i} \in \bool$, $\forall a \in \intent{1}{\bNA}$, $i \in \left\{l, r\right\}$: sign of $i$ input of adder $a$. $0$ for $+$ and $1$ for $-$; \\
		$c_{a,i,k} \in \bool$, $\forall a \in \intent{1}{\bNA}$, $i \in \left\{l, r\right\}$, $k \in \intent{0}{\bNA-1}$: $1$ if input $i$ of adder $a$ is adder $k$; \\
		$\sigma_{a,s} \in \bool$, $\forall a \in \intent{1}{\bNA}$, $s \in \intent{0}{\Smax}$: $1$ if left shift before adder $a$ is equal to $s$; \\
		$\Psi_{a, s} \in \bool$, $\forall a \in \intent{1}{\bNA}$, $s \in \intent{\Smin}{0}$: $1$ if negative shift of adder $a$ is equal to $s$; \\
		$o_{a,j} \in \bool$, $\forall a \in \intent{1}{\bNA}$, $j \in \intent{1}{N_O}$: $1$ if adder $a$ is equal to the $j$-th target constant. \\
		\bottomrule
	\end{tabularx}
	\caption{Constants (top) and variables (bottom) used in the ILP formulation}\label{tab:mcm_data_var}
\end{table}

Equation~\eqref{eq:newadderrel} is nonlinear, thus multiple intermediate variables are necessary to compute $c_a$ using linear constraints only.
For instance, in order to linearize
\begin{equation}\label{eq:linearizeshifts}
	c^{\sh}_{a,l} = 2^{s_{a,l}} c_{a,l},
\end{equation}
which involves a power of two and a product, a set of binary variables $\sigma_{a,s}$ is required with $s$ taking values ranging from $0$, meaning no shift, to an upper bound on the possible shift $\Smax$.
These variables are used in \eqref{cst:model1_valueshifted} where a constant, $2^0, 2^1, 2^2$, etc., is multiplied with a variable removing the power of two and product of variables issues.
Variable $\sigma_{a,s}$ serves as an indicator of the shift $s$ being used for adder $a$, $\forall s \in \intent{0}{\Smax}$, and only one shift is selected for each adder.
All the required intermediate variables are presented in \tablename~\ref{tab:mcm_data_var}.
These variables are used to produce the following linear model describing the adder graph topology and the link between fundamentals and target constants $C_j$:\par\nobreak
{%
\newcount\tmptheequation
\tmptheequation=\theequation
\renewcommand{\theequation}{C\arabic{equation}}
\setcounter{equation}{0}
\par\begingroup\hsize=1.08\linewidth
\allowdisplaybreaks
\leqnomode
{
\small
\begin{fleqn}[8mm]
\begin{align}
& c^\nsh_a = c_{a,l}^{\sh, \sg} + c_{a,r}^{\sh, \sg} && \forall a \in \intent{1}{N_A} \label{cst:model1_addervaluesum} \\
& c_a^{\nsh} = 2^{-s} c_a \quad \text{ if } \Psi_{a,s} = 1 && \forall a \in \intent{1}{N_A}, s \in \intent{\Smin}{0} \label{cst:model1_addervalue} \\
& \sum\limits_{s=\Smin}^{0}{\Psi_{a,s}} = 1 && \forall a \in \intent{1}{N_A} \label{cst:model1_negshift} \\
& \sigma_{a,0} = \sum\limits_{s=\Smin}^{-1}{\Psi_{a,s}} && \forall a \in \intent{1}{N_A} \label{cst:model1_onlynegshiftifnopos} \\
& c_a = 2c_a^\odd + 1 && \forall a \in \intent{1}{N_A} \label{cst:model1_odd} \\
& c_{a,i} = c_k \quad \text{ if } c_{a,i,k} = 1 && \forall a \in \intent{1}{N_A}, i \in \left\{l, r\right\}, \label{cst:model1_previousforcevalue}\\ &&&\forall k \in \intent{0}{a-1} \nonumber\\
& \sum\limits_{k=0}^{a-1}{c_{a,i,k}} = 1 && \forall a \in \intent{1}{N_A}, i \in \left\{l, r\right\} \label{cst:model1_onlyoneinput} \\
& c_{a,l}^{\sh} = 2^s c_{a,l} \quad \text{ if } \sigma_{a,s} = 1 && \forall a \in \intent{1}{N_A}, s \in \intent{0}{\Smax} \label{cst:model1_valueshifted} \\
& \sum\limits_{s=0}^{\Smax}{\sigma_{a,s}} = 1 && \forall a \in \intent{1}{N_A} \label{cst:model1_onlyoneshift} \\
& c_{a,i}^{\sh, \sg} = -c_{a,i}^{\sh} \quad \text{ if } \Phi_{a,i} = 1 && \forall a \in \intent{1}{N_A}, i \in \left\{l, r\right\} \label{cst:model1_signneg} \\
& c_{a,i}^{\sh, \sg} = c_{a,i}^{\sh} \quad \text{ if } \Phi_{a,i} = 0 && \forall a \in \intent{1}{N_A}, i \in \left\{l, r\right\} \label{cst:model1_signpos} \\
& c_a = C_j \quad \text{ if } o_{a,j} = 1 && \forall a \in \intent{0}{N_A}, j \in \intent{1}{N_O} \label{cst:model1_endingconstantforcevalue} \\
& \sum\limits_{a=0}^{N_A}{o_{a,j}} = 1 && \forall j \in \intent{1}{N_O} \label{cst:model1_eachconstanthasanadder}
\end{align}
\end{fleqn}
}%
\reqnomode
\par\endgroup
\renewcommand{\theequation}{\arabic{equation}}
\setcounter{equation}{\tmptheequation}
}%

In details, constraint~\eqref{cst:model1_addervaluesum} states that the value of an adder before a potential negative shift is equal to the sum of its shifted and signed inputs.
Constraints~\eqref{cst:model1_addervalue}, \eqref{cst:model1_negshift} and \eqref{cst:model1_odd} apply the negative shift to variables $c_a$ and ensure that the computed fundamental is odd.
It can be noticed that a potential negative shift after $c_a^{\nsh}$ only makes sense if the sum of the inputs is even, which can only happen if no left shift is applied to the left input: constraint~\eqref{cst:model1_onlynegshiftifnopos} specifically states this in order to speed up the solving process.
The link between an adder and its inputs is enforced by constraints~\eqref{cst:model1_previousforcevalue} and \eqref{cst:model1_onlyoneinput}.
Constraints~\eqref{cst:model1_valueshifted} and \eqref{cst:model1_onlyoneshift} permit to apply the shift to the left input of an adder while constraints \eqref{cst:model1_signneg} and \eqref{cst:model1_signpos} apply the sign.
Finally, every target constant $C_j$ is computed once and only once thanks to constraints \eqref{cst:model1_endingconstantforcevalue} and \eqref{cst:model1_eachconstanthasanadder}.

These constraints are sufficient to define a model for the following decision problem: is there an adder graph with $\bNA$ adders where all the target constants $C_j$ are computed as fundamentals?
Either the model is ``infeasible'', meaning that no adder graph exists for computing the target constants with just $\bNA$ adders, or ``optimal'' returning values $c_a$, that encode fundamentals, and corresponding shifts and signs that permit to produce a valid adder graph as in \figurename~\ref{fig:compareonebit}.

To tackle \MCM as minimization problem, which is desirable in order to use the full potential of MILP solvers, $\bNA$ can be fixed to a known upper bound, obtained using a heuristic solution~\cite{VoronenkoPueschel_Multiplierlessmultipleconstant_2007, KummZipfFaustChang_Pipelinedaddergraph_2012} or a greedy algorithm~\cite{Bernstein_Multiplicationintegerconstants_1986}, and then the objective function is to minimize the number of \emph{effectively used} adders.
A binary variable, $u_a \in \bool$, $\forall a \in \intent{1}{\bNA}$, permits to deactivate an adder if not used: $c_a = 0 \text{ if } u_a = 0$.
Then, adding to the model the objective function
\begin{equation}\label{eq:mcmobjective}
	\min \sum u_a,
\end{equation}
permits to solve the MCM problem as a minimization problem.

\subsection{Bounding and/or Minimizing Adder Depth}\label{sec:mcmbmcm}

The latency of the circuit is directly related to the number of cascaded adders or adder depth (AD).
It is often preferable to have a bound on the AD to ensure a bound on the latency, or at least to have the minimal AD possible as a second objective, while simultaneously minimizing the number of adders.
Works \cite{Gustafsson_Towardsoptimalmultiple_2008,Kumm_MultipleConstantMultiplication_2016_book} that rely on the search space enumeration solve the \MCM problem with an AD bound up to $3$ or $4$ in best case.
In this section we propose a new simple way to propagate the AD and to optionally bound it by a user-given constant $\overline{\ad}$ and/or to minimize it as a second objective.

In order to track the AD, we introduce the variable $\ad_{\max}$, and two sets of integer variables: $\ad_a$ and $\ad_{a,i}$, $\forall a \in \intent{1}{\bNA}$, $i \in \left\{l,r\right\}$, representing the AD of the adder $a$ and the AD of its left and right inputs, respectively.
Naturally, the adder depth of the input is set at zero, $\ad_{0} = 0$, and the AD propagation and the bound are handled by the following constraints:
\begin{align}
& \ad_a = \max\left(\ad_{a,l} + 1, \ad_{a,r} + 1\right), && \forall a \in \intent{1}{\bNA}, \label{eq:adderdepthpropagation} \\
& \ad_{a,i} = \ad_k \quad \text{ if } c_{a,i,k} = 1, && \forall a \in \intent{1}{\bNA}, i \in \left\{l, r\right\}, \nonumber\\ &&& \forall k \in \intent{0}{a-1}, \label{eq:adpropagation} \\
& \ad_{\max} \geq \ad_a, && \forall a \in \intent{1}{\bNA}, \label{eq:adderdepthvalue} \\
& \ad_{\max} \leq \overline{\ad}. \label{eq:adderdepthbound} &&
\end{align}
Note that the $\max$ in \eqref{eq:adderdepthpropagation} can be linearized adding a set of binary variables to the model.
Constraint \eqref{eq:adderdepthvalue} ensures that the integer variable $\ad_{\max}$ will be equal or greater than actual adder depth.
With \eqref{eq:adderdepthbound}, we guarantee that the adder depth is bounded by $\overline{\ad}$.
In contrast to existing approaches \cite{Gustafsson_Towardsoptimalmultiple_2008} and \cite{Kumm_MultipleConstantMultiplication_2016_book}, this encoding of AD is performed automatically by the solver, and does not require precomputations or a large number of variables and constraints.

We further extend our ILP model towards the bi-objective problem \MCMad in order to optimize for both adder count and AD.
The variable $\ad_{\max}$, which encodes the AD of the adder graph, is used to replace the objective function \eqref{eq:mcmobjective} by
\begin{equation}\label{eq:mcmadobjective}
	\min \sum_{a=1}^{\bNA}{\left(\bNA u_a\right)} + \ad_{\max}.
\end{equation}
This new objective function is a weighted sum that enforces a lexicographic optimization with the number of adders as first objective and the AD as second.
Indeed, reducing the number of adders is unconditionally stronger than increasing the AD because $\bNA \geq \mathit{ad}_{\max}$.

Solving \MCMad permits to select from the set of solutions with minimal number of adders those which yield the smallest delay in a hardware implementation.
For example, in \figurename~\ref{fig:compareonebit}, we propose two optimal solutions in terms of the number of adders, $\NA = 3$, for the target constants $\left\{49,51\right\}$, but different adder depths: $\AD = 3$ in \figurename~\ref{fig:4951high} and $\AD = 2$ in \figurename~\ref{fig:4951low}.
Solving \MCMad would directly find the adder graph with the lowest adder depth among the optimal ones in terms of number of adders.

This problem can be difficult for the solver, thus, in order to speed up the solving process, we provide redundant constraints that should help to reduce the search space.
Gustafsson~\cite{Gustafsson_LowerBoundsConstant_2007} proposed some lower bounds on the AD for the target constants which we can use in the following way:
\begin{align}
& \ad_a \geq o_{a,j} \times \underline{\ad}_j && \forall a \in \intent{1}{\bNA}, j \in \intent{1}{N_O}, & \label{eq:adderdepthoutputbound}
\end{align}
where $\underline{\ad}_j$ is the lower bound on the adder depth of the target constant indexed by $j$.
For our $\left\{49,51\right\}$ target set, the first constant $49$ has a minimal adder depth $\underline{\ad}_1 = 2$, hence the constraint becomes:
\begin{align}
	& \ad_a \geq 2 o_{a,1}, && \forall a \in \intent{1}{\bNA},
\end{align}
enforcing the adder $a$ to have an AD of at least $2$ before considering it as a potential output for the first target constant~$49$.

In Section~\ref{sec:expe}, we will show the advantages of \MCMad over the classic single-objective \MCMA. Optimal solutions are obtained in reasonable time when solving either problems, and solving \MCMad provides better solutions.
For that reason, we will also consider the AD minimization as second objective when tackling \MCMB and \TMCM.

Similarly to bounding and/or minimizing the adder depth, the \emph{Glitch Path Count} (GPC) metric could be targeted in order to control the power consumption~\cite{DemirsoyDempsterKale_Poweranalysismultiplier_2002}. In~\cite{Kumm_OptimalConstantMultiplication_2018}, Kumm showed how to include such metric into an ILP model and this can be adopted for our solution.

\section{Optimal Multiple Constant Multiplication Counting One-Bit Adders}\label{sec:mcm1b}

Minimizing the number of adders and the AD leads to efficient hardware but can also be used in software.
In many cases, optimizing \wrt{} these metrics is the best we can provide and they already lead to a significant cost reduction.
However, when specifically targeting hardware, \eg{}, Application-Specific Integrated Circuits (ASICs) or Field Programmable Gate Arrays (FPGAs), and when the word length of the input is \emph{a priori} known, it is possible to optimize using a finer-grain metric: the number of one-bit adders.
Indeed, two optimal adder graphs, \wrt{} the number of adders, can have a significantly different number of one-bit adders.
This is illustrated by \figurename~\ref{fig:compareonebit}: for a 3-bit input $x$, solution in \figurename~\ref{fig:4951high} requires $22$ one-bit adders, while the one in \figurename~\ref{fig:4951low} needs only $9$ one-bit adders.
These large differences motivate the choice to specifically target the one-bit adder metric.

To solve the \MCMB problem, we enhance our \MCMad model.
In addition to the constraints fixing the adder graph topology and keeping track of the adder depth, we need to be able to count the number of one-bit adders.
This additional need comes with two main difficulties: first, we have to be able to propagate the data word length which directly impacts the number of one-bit adders required for a given adder; second, counting the number of one-bit adders requires to consider many cases, as presented in \cite{JohanssonGustafssonWanhammar_BitLevelOptimization_2007}, \eg{}, the cost in terms of one-bit adders differs between two adders because of shifts and subtractions.

The former problem of word length propagation can be decomposed into tracking the Most Significant Bit (MSB) and Least Significant Bit (LSB) positions.
Integer additions and subtractions impact the MSB but do not change the LSB, hence without loss of generality we can consider that the LSB is always equal to zero.
Given an upper bound on the input, $\overline{x}$, it is possible to compute the required MSB after any adder:
\begin{equation}\label{eq:msbcomputebase}
	\msb_a = \left\lceil\log_2\left(\overline{x} c_a\right)\right\rceil
\end{equation}
where $c_a$ is the adder fundamental.
Incorporating this nonlinear constraint into our ILP model requires a few adjustments.
First, the rounding can be removed by relaxing the equality. Second, exponentiation of both sides to remove the $\log_2$ leads~to
\begin{equation}\label{eq:msbcompute}
2^{\msb_a} \geq \overline{x} c_a.
\end{equation}
The expression $2^{\msb_a}$ can be linearized similarly to shifts with the challenge of the wider range of possible values for $\msb_a$.
This permits to propagate the data word length, which is, in the worst case, equal to the number of one-bit adders, $B_a$:
\begin{equation}
	B_a \leq \msb_a + 1.
\end{equation}
However, multiple corner cases permit to save one-bit adders.

\begin{figure}
	\centering
	\begin{subfigure}{0.43\linewidth}
		\centering
		\includegraphics[width=1.0\linewidth]{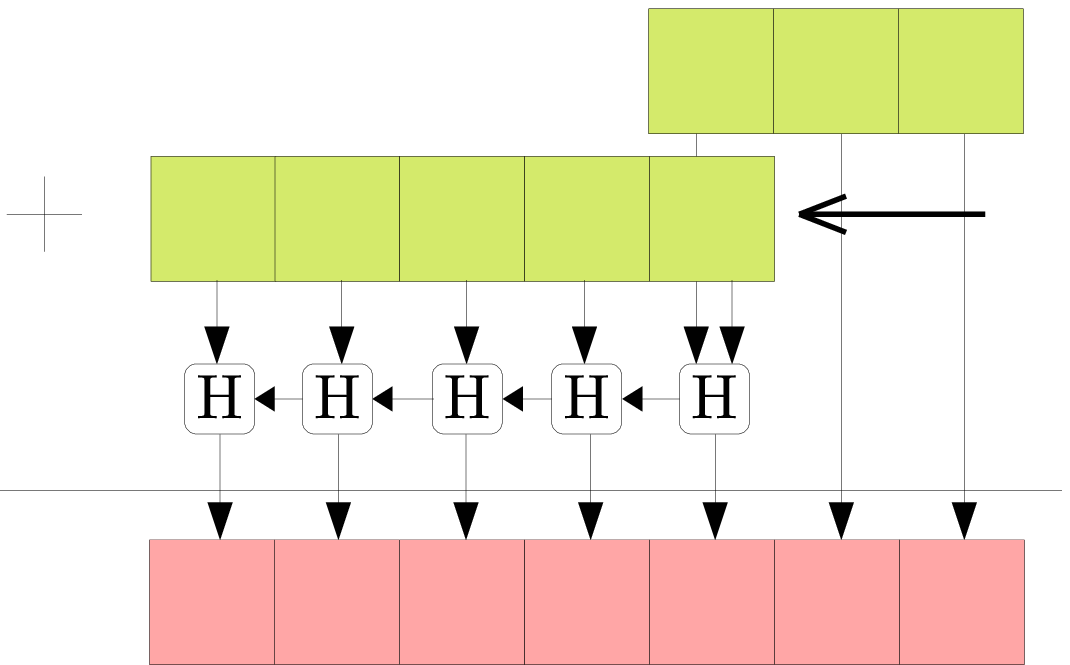}
		\caption{Some output bits are computed without one-bit adders thanks to shifts.}\label{fig:onebitaddershift}
	\end{subfigure}
	\hspace{0.08\linewidth}
	\begin{subfigure}{0.43\linewidth}
		\centering
		\includegraphics[width=1.0\linewidth]{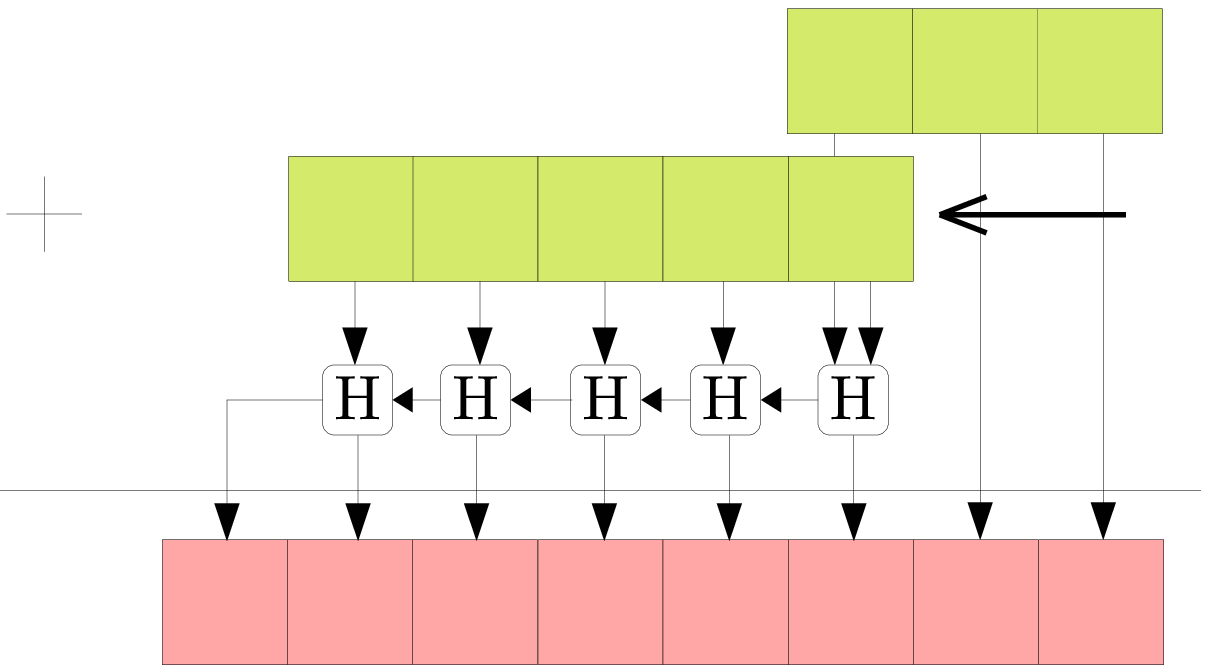}
		\caption{In some cases, the MSB can be computed by the carry of the last one-bit adder.}\label{fig:onebitaddercarryshift}
	\end{subfigure}

	\medskip

	\begin{subfigure}{0.43\linewidth}
		\centering
		\includegraphics[width=1.0\linewidth]{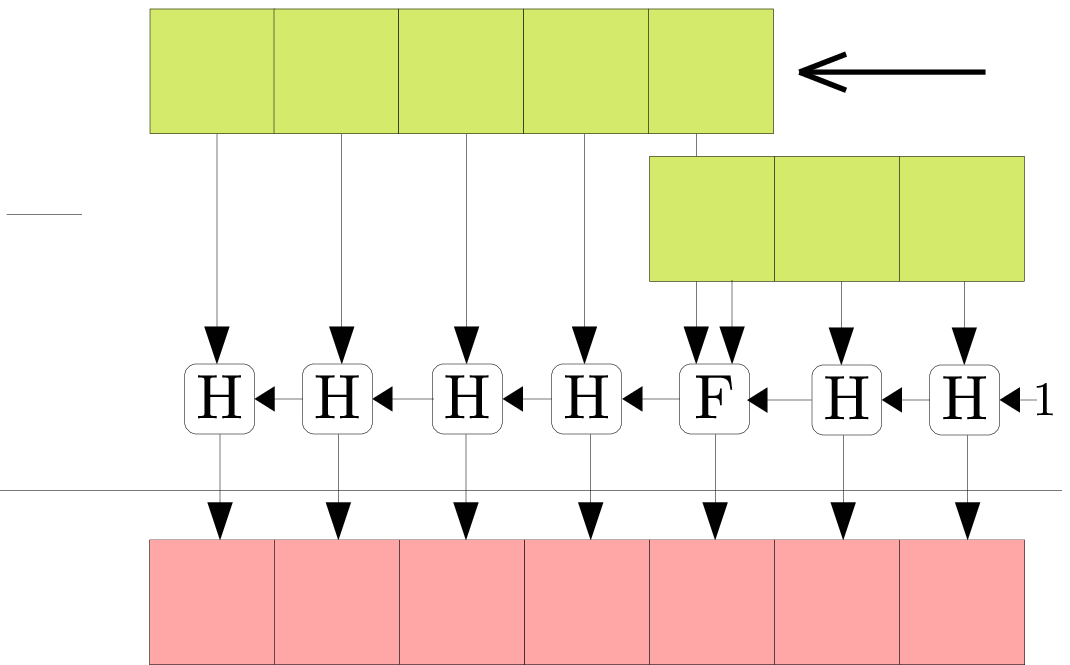}
		\caption{For subtraction, shifts do not always permit to save one-bit adders.}\label{fig:onebitaddersubtract}
	\end{subfigure}
	\hspace{0.08\linewidth}
	\begin{subfigure}{0.43\linewidth}
		\centering
		\includegraphics[width=1.0\linewidth]{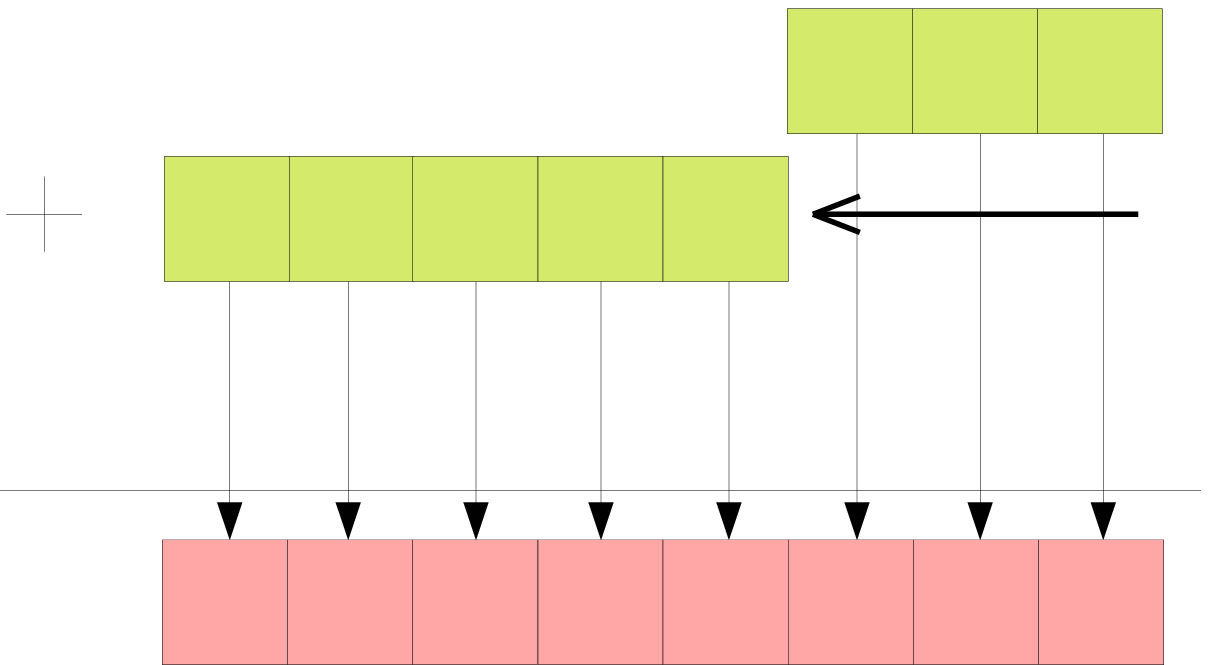}
		\caption{When shifts are large enough, no one bit adders are required.}\label{fig:onebitaddernone}
	\end{subfigure}
	\caption{Counting one-bit adders}\label{fig:onebitaddersimplecases}
\end{figure}

As illustrated in \figurename{}~\ref{fig:onebitaddershift}, shifts can lead to output bits computed without one-bit adders.
Yet, one should note that it is not always the case with subtractions, see \figurename{}~\ref{fig:onebitaddersubtract}.
The MSB computation is either done by a dedicated one-bit adder, as represented in \figurename{}~\ref{fig:onebitaddershift}, or obtained from the carry of the last one-bit adder, as in \figurename{}~\ref{fig:onebitaddercarryshift}.
This can be deduced from the MSB positions which depend on the fundamental values, see \eqref{eq:msbcomputebase}, allowing for the gain of a one-bit adder.
One-bit adders might even not be necessary at all, as illustrated in \figurename{}~\ref{fig:onebitaddernone}.

To wrap up, counting the number of one-bit adders, $\OB_a$, consists in computing the word length and deducting the gains $g_a$ and $\psi_a$,
\begin{equation}
\OB_a = \msb_a + 1 - g_a - \psi_a,
\end{equation}
where $\psi_a$ is equal to one if the MSB is computed by the carry of the last one-bit adder.
The gain $g_a$ is dependent on the signs, shifts and MSBs and can be summarized as follows:
\begin{equation}\label{eq:gainob}
g_a = \left\{\begin{array}{ll}
\msb_a & \text{if } s_{a,l} > \msb_{a,r} \\
s_{a,l} & \text{if } \Phi_{a,r} = 0 \\
0 & \text{otherwise}
\end{array}\right.,
\end{equation}
formalizing cases illustrated in \figurename~\ref{fig:onebitaddershift}, \ref{fig:onebitaddersubtract} and \ref{fig:onebitaddernone}.
To include the computation of $\OB_a$ and $g_a$ in an ILP model, we add several binary variables to handle each condition.

As each adder $a$ has an associated cost $B_a$, it is straightforward to formulate the objective as:
\begin{equation}\label{eq:obobjfun}
	\min \sum_{a=1}^{\bNA}{\bNA\OB_a} + \ad_{\max},
\end{equation}
where $\bNA$ is the number of adders the adder graph can use.
This solves the bi-objective problem minimizing the number of one-bit adders first and the adder depth second.

For a given instance, we make the following conjecture:
if a solution of that instance requires $\NA$ adders, then the optimal solution in terms of number of one-bit adders also requires at most $\NA$ adders.
As a consequence, we will use a bound on $\bNA$ and assume we do not exclude optimal solutions with respect to one-bit adders.

In Section~\ref{sec:expe}, our benchmarks show that solving \MCMB, on average, reduces the number of LUTs by $\gainLUTsOBAvsNA\%$ compared to our \MCMA.

\section{Truncated Multiple Constant Multiplication}\label{sec:tmcm1b}

\begin{figure}
	\centering
	\includegraphics[width=0.75\linewidth]{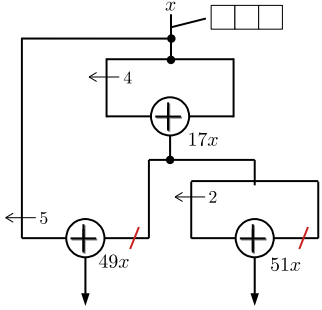}
	\caption{Computing $49x$ and $51x$ via the fundamental $17$ leads to only $4$ one-bit adders for a 3-bit input/output.
	}\label{fig:4951trunc}
\end{figure}

\begin{figure}
	\centering
	\begin{subfigure}{0.43\linewidth}
		\centering
		\begin{tikzpicture}
	\small
	\hspace{-7mm}
	\matrix (adderclassic) [row sep=0mm, column sep=-14mm]
	{
		\node[dspnodeopen,dsp/label=above, label={below:$\phantom{s_{a,l}}\longleftarrow s_{a,l}$}] (m11) {$\left(\varepsilon^{\inf}_{a,l}, \varepsilon^{\sup}_{a,l}\right)$}; \pgfmatrixnextcell
		\node[coordinate] (m12) {}; \pgfmatrixnextcell
		\node[dspnodeopen,dsp/label=above, label={below:$\phantom{s_{a,r}\longleftarrow s_{a,r}}$}] (m13) {$\left(\varepsilon^{\inf}_{a,r}, \varepsilon^{\sup}_{a,r}\right)$}; \\
		\node[coordinate] (m21) {}; \pgfmatrixnextcell
		\node[dspadder, label={above right:$\phantom{\phi_{a,r}}$}, label={above left:$\phantom{\phi_{a,l}}$}, label={above right:$t_{a,r}$}, label={above left:$t_{a,l}$}, label={right:$\ /$}, label={left:$/\ $}, label={below right:$\left(\varepsilon^{\inf}_a, \varepsilon^{\sup}_a\right)$}, label={below left:\phantom{$\left(\varepsilon^{\inf}_a, \varepsilon^{\sup}_a\right)$}}] (m22) {}; \pgfmatrixnextcell
		\node[coordinate] (m23) {}; \\
		\node[coordinate] (m31) {}; \pgfmatrixnextcell
		\node[dspnodeopen] (m32) {}; \pgfmatrixnextcell
		\node[coordinate] (m33) {}; \\
	};
	\begin{scope}[start chain]
		\chainin (m11);
		\chainin (m21) [join=by dspline];
		\chainin (m22) [join=by dspconn];
	\end{scope}
	\begin{scope}[start chain]
		\chainin (m13);
		\chainin (m23) [join=by dspline];
		\chainin (m22) [join=by dspconn];
	\end{scope}
	\begin{scope}[start chain]
		\chainin (m22);
		\chainin (m32) [join=by dspconn];
	\end{scope}
\end{tikzpicture}
		\vspace{-1em}
		\caption{Model of an adder in presence of errors $\varepsilon$ and truncations $t$.}\label{fig:ilpmodelerrortruncate}
	\end{subfigure}
	\hspace{0.08\linewidth}
	\begin{subfigure}{0.43\linewidth}
		\centering
		\includegraphics[width=1.0\linewidth]{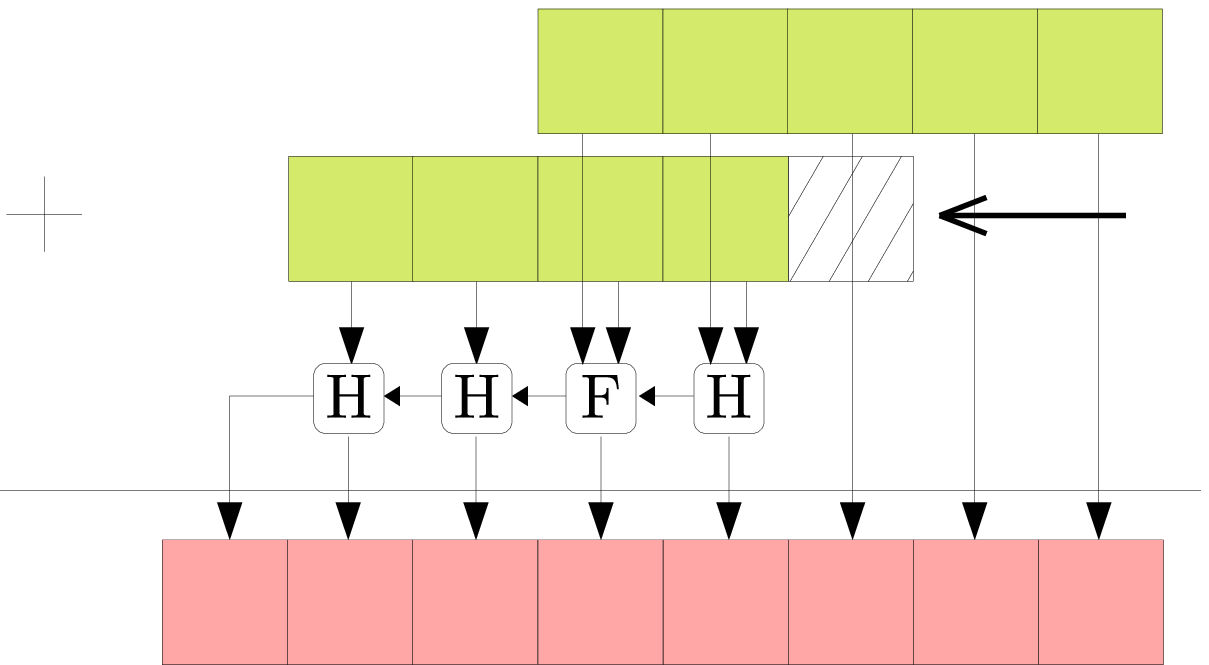}
		\caption{Some output bits are computed without one-bit adders thanks to shifts and truncations.}\label{fig:onebitaddertruncate}
	\end{subfigure}

	\medskip

	\begin{subfigure}{0.43\linewidth}
		\centering
		\includegraphics[width=1.0\linewidth]{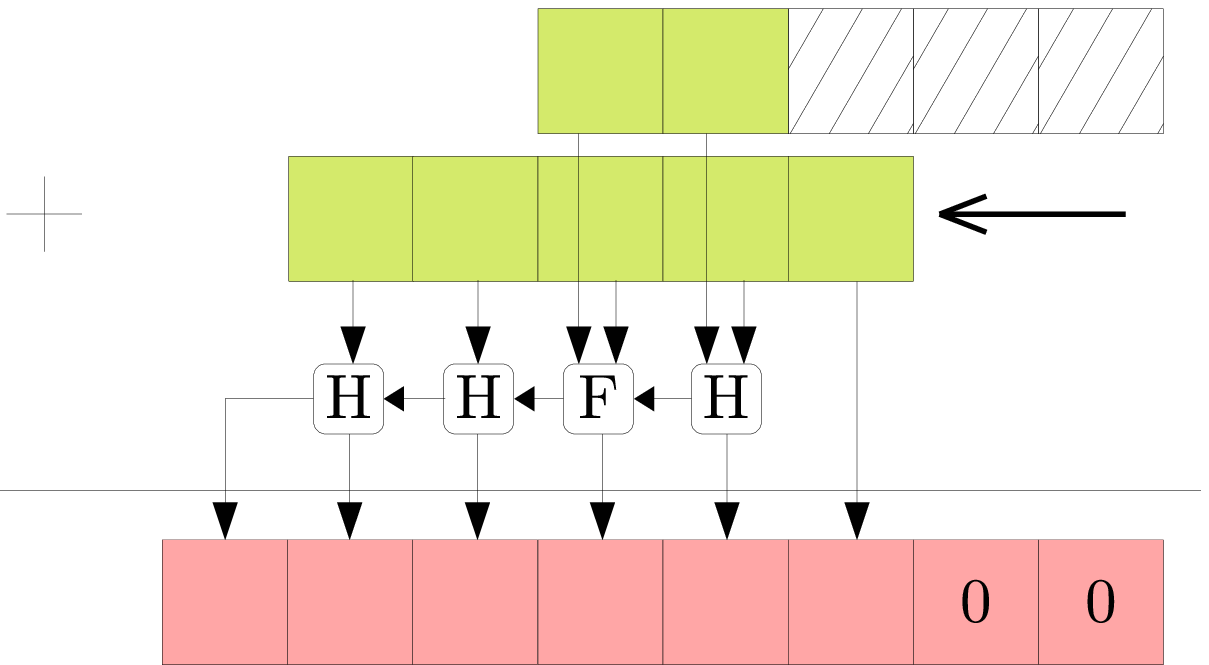}
		\caption{Truncations might induce zeros in the output.}\label{fig:onebitaddertruncatefreefirst}
	\end{subfigure}
	\hspace{0.08\linewidth}
	\begin{subfigure}{0.43\linewidth}
		\centering
		\includegraphics[width=1.0\linewidth]{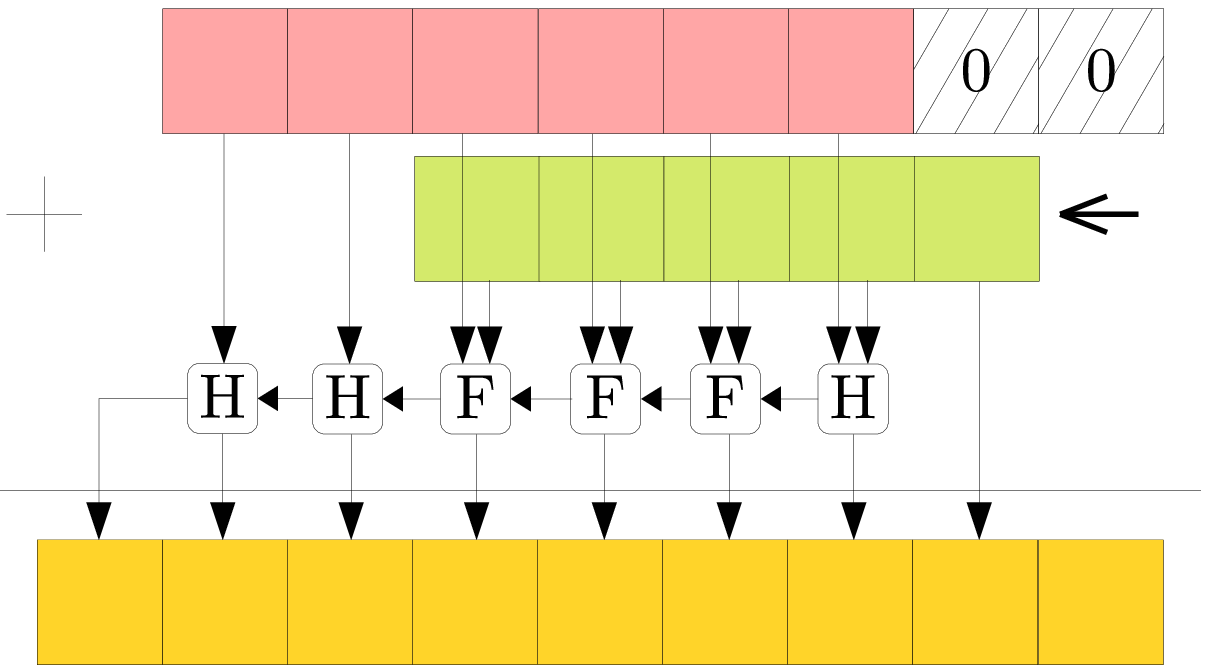}
		\caption{Zeros can be truncated without increasing the error.}\label{fig:onebitaddertruncatefreesecond}
	\end{subfigure}
	\caption{Counting one-bit adders and propagating errors in presence of truncations}\label{fig:onebitadderstrunc}
\end{figure}

Arithmetic operation usually increase the size of the data paths but the full-precision result is often not required. For instance, consider recursive digital filters, in which the result of MCM is typically rounded to some internal word length at each iteration. In these cases, hardware designers can often provide an absolute error bound, $\overline{\varepsilon}$, that the computed products should respect. In order to avoid wasting resources to compute the bits that will not impact the rounded result, we propose to introduce truncations inside the adder graphs. \figurename~\ref{fig:onebitaddertruncate} illustrates a case when truncating a bit in one of the adder's operands saves a one-bit adder. Of course, a truncation can generate an error that should be bounded and propagated, and the truncated adder graph should respect the bound $\overline{\varepsilon}$.
Incorporating truncations the complexity of the problem, however it has been shown that it can significantly reduce the final hardware cost \cite{GuoDeBrunnerJohansson_TruncatedMCMusing_2010, DinechinFilipKummForget_TableBasedversus_2019}.

In contrast to previous works, which focus on truncations of a \textit{fixed} adder graph, we postulate that truncations should guide the topology, potentially leading to a different adder graph than the one obtained solving the \MCMB problem. Consider for instance the multiplication by $49$ and $51$ in \figurename~\ref{fig:compareonebit}, where the adder graph \figurename~\ref{fig:4951low} is the \MCMB solution. Applying the truncations, s.t. only 3 output bits are faithful, yields a design requiring at least 5 one-bit adders. However, there exists another adder graph topology, shown in~\figurename~\ref{fig:4951trunc}, that requires only 4 one-bit adders and is hence better suited for a target truncated result.

 In this Section we outline the ILP model permitting to solve the \TMCM as \textit{one} optimization problem, covering the whole design space. In Section~\ref{sec:expe}, our experiments demonstrate that when the goal is to keep only half of the output bits, our \TMCM solutions reduce the number of LUTs by $\gainLUTsThalfvsOBA\%$ compared to our new \MCMB, and by $\gainLUTsThalfvsNA\%$ compared to the state of the art \MCMA, on average.

\subsection{Modeling truncations and error propagation}

Truncations permit to gain one-bit adders and should be incorporated into a model in order to benefit from this possibility.
To do so, we add integer variables, $t_{a,i}$, encoding the position of the truncated bits from the left and right inputs of adder $a$.
Then, we update \eqref{eq:gainob} in order to include the truncations in the one-bit adders gain count:
\begin{equation}\label{eq:gainobtrunc}
g_a = \left\{\begin{array}{ll}
\msb_a & \text{if } s_{a,l} > \msb_{a,r}, \\
\max\left(t_{a,l}, s_{a,l}, t_{a,r}\right) & \text{if } s_{a,l} \geq 0 \land \Phi_{a,l} = \Phi_{a,r}, \\
\max\left(t_{a,l}, s_{a,l}\right) & \text{if } s_{a,l} \geq 0 \land \Phi_{a,l} = 1, \\
t_{a,r} & \text{if } s_{a,l} \geq 0 \land \Phi_{a,r} = 1, \\
0 & \text{otherwise.}
\end{array}\right.
\end{equation}
Typically, this gain is maximized in order to decrease the number of one-bit adders in the adder graph.
Obviously, truncations should be constrained by the error they induce.
These errors must be propagated and respect a user-given bound for each output.

Previous works \cite{DinechinFilipKummForget_TableBasedversus_2019, GarciaVolkovaKumm_TruncatedMultipleConstant_2022} underestimate the truncation-induced errors resulting in an output error exceeding the bound $\overline{\varepsilon}$.
This underestimation was partly mitigated by an overestimation of the propagation of errors that we discuss~here.

Classically, in computer arithmetic we consider the following model to link the exact quantities and errors:
\begin{equation}\label{eq:classicerrorformula}
	\widetilde{y} = y + \Delta, \quad \text{where } \left|\Delta\right| \leq \varepsilon,
\end{equation}
where $\widetilde{y}$ is the actually computed quantity, $y$ is the exact quantity, and $\varepsilon$ is a bound on the error.
This fits well for symmetric errors, which is the case when rounding to nearest, but overestimates nonsymmetric truncation-induced errors.
We propose to track errors using two positive bounds, $\varepsilon^{\inf}$ and $\varepsilon^{\sup}$, the former for the negative deviation from the exact quantity and the latter for the positive deviation.
That way, it is possible to closely link the computed quantity with the exact quantity:
\begin{equation}\label{eq:newerrorformula}
	y - \varepsilon^{\inf} \leq \widetilde{y} \leq y + \varepsilon^{\sup}.
\end{equation}
This equation holds for positive quantities and the following error-analysis assume that all the quantities are positives.
Minor sign adjustments of $\varepsilon$'s and reversing inequality relations are necessary to extend our analysis to negative quantities.

In order to correctly propagate the errors, we start by investigating the evolution of errors when applying unary operators used in truncated adder graphs.
These unary operators are the shift, the negation and the truncation.
First, applying a shift $s$ to an inexact quantity $\widetilde{y}$ is straightforward and as it increases both error bounds:
\begin{equation}\label{eq:shifterrorformula}
2^s y - 2^s \varepsilon^{\inf} \leq \widetilde{2^s y} \leq 2^s y + 2^s \varepsilon^{\sup}.
\end{equation}
Second, the negation applied on an inexact quantity $\widetilde{y}$ does not increase the overall error but swaps the deviations from the exact quantity $y$ like so:
\begin{equation}\label{eq:negerrorformula}
- y - \varepsilon^{\sup} \leq -\widetilde{y} \leq - y + \varepsilon^{\inf}.
\end{equation}

Finally, when applied to a quantity $\widetilde{y}$, truncation up to the $t$-th bit, $\diamond_t\left(\cdot\right)$, removes information.
This can increase the negative deviation $\varepsilon^{\inf}$ but not the positive deviation $\varepsilon^{\sup}$, thus the error bounds increase asymmetrically:
\begin{equation}
	y - \varepsilon^{\inf} - \varepsilon_t \leq \diamond_t\!\left(\widetilde{y}\right) \leq y + \varepsilon^{\sup} + 0,
\end{equation}
where $\varepsilon_t$ is bounded by the quantity it removes from $\widetilde{y}$:
\begin{equation}\label{eq:truncerrorbound}
\varepsilon_t \leq 2^{t} - 1.
\end{equation}
The bound is reached when all truncated bits are $1$'s.
However, truncations could induce bits equal to $0$ in the data path, as illustrated in \figurename~\ref{fig:onebitaddertruncatefreefirst}, and keeping track of these trailing $0$'s, denoted $z$, allows for tighter bound on the truncation errors:
\begin{equation}\label{eq:truncerrorboundzeros}
\varepsilon_t \leq \max\left(2^t - 2^z, 0\right).
\end{equation}
To be able to use this bound, we add an integer variable $z_a$, for each adder $a$, that keeps track of the number of trailing $0$'s induced by truncations or propagated from the inputs.
In some cases where truncated bits are all zeros, as in \figurename~\ref{fig:onebitaddertruncatefreesecond}, we can even have error-free truncations.
And with our approach, the ILP solver will automatically privilege those.

Finally, we need to propagate errors through adders which, for two inexact inputs, $\widetilde{y_1}$ and $\widetilde{y_2}$, adds error bounds together:
\begin{align}
y_1 + y_2 - \varepsilon^{\inf}_{y_2} - \varepsilon^{\inf}_{y_2} \leq \widetilde{y_1}+\widetilde{y_2} \leq y_1 + y_2 + \varepsilon^{\sup}_{y_1} + \varepsilon^{\sup}_{y_2}
\end{align}

For each operand, we defined a dedicated propagation rule.
Our end goal is to bring them all and to bind them in a single error propagation rule.
For each adder, $a$, we define variables for error bounds $\left(\varepsilon^{\inf}_a, \varepsilon^{\sup}_a\right)$ which need to be bounded by the user-given acceptable output error $\overline{\varepsilon}$:
\begin{equation}
\left|\varepsilon^{\inf}_a\right| \leq \overline{\varepsilon} \quad \text{ and } \quad \left|\varepsilon^{\sup}_a\right| \leq \overline{\varepsilon}.
\end{equation}
Together with left and right truncations $t_{a,i}$, we propagate $\left(\varepsilon^{\inf}_{a,i}, \varepsilon^{\sup}_{a,i}\right)$, and \figurename~\ref{fig:ilpmodelerrortruncate} sums up the interconnections between these variables.

The propagation rule for each adder $a$ is as follows:
\begin{align}
	& \varepsilon^{\inf}_a = \left\{\hspace{-2mm}\begin{array}{ll}
	\left(2^{s_{a,l}}\varepsilon^{\inf}_{a,l} + \varepsilon_{t_{a,l}}\right) + \left(\varepsilon^{\inf}_{a,r} + \varepsilon_{t_{a,r}}\right), & \hspace{-3mm} \text{if } \Phi_{a,l} = \Phi_{a,r}, \\
	\left(2^{s_{a,l}}\varepsilon^{\inf}_{a,l} + \varepsilon_{t_{a,l}}\right) + \left(\varepsilon^{\sup}_{a,r} + \varepsilon_{t_{a,r}}\right), & \hspace{-3mm} \text{if } \Phi_{a,r} = 1, \\
	\left(2^{s_{a,l}}\varepsilon^{\sup}_{a,l} + \varepsilon_{t_{a,l}}\right) + \left(\varepsilon^{\inf}_{a,r} + \varepsilon_{t_{a,r}}\right), & \hspace{-3mm} \text{if } \Phi_{a,l} = 1,
	\end{array}\right. \\
	& \varepsilon^{\sup}_a = \left\{\hspace{-2mm}\begin{array}{ll}
	2^{s_{a,l}}\varepsilon^{\sup}_{a,l} + \varepsilon^{\sup}_{a,r}, & \text{if } \Phi_{a,l} = \Phi_{a,r}, \\
	2^{s_{a,l}}\varepsilon^{\sup}_{a,l} + \varepsilon^{\inf}_{a,r}, & \text{if } \Phi_{a,r} = 1, \\
	2^{s_{a,l}}\varepsilon^{\inf}_{a,l} + \varepsilon^{\sup}_{a,r}, & \text{if } \Phi_{a,l} = 1.
	\end{array}\right.
\end{align}

Finally, the propagated errors can force the result of the adder to the next binade \wrt{} the one deduced by \eqref{eq:msbcompute}.
To prevent the risk of overflow, we adjust the MSB taking the error into account:
\begin{equation}\label{eq:msbcomputeerror}
	2^{\msb_a} \geq \overline{x} c_a + \left|\varepsilon^{\sup}_a\right|,
\end{equation}
where $\overline{x}$ is an upper bound on the maximum of the absolute values of lower and upper bounds of $x$.
The solver will chose itself if it is more interesting to increase the MSB in order to gain more from truncations or not.

We do not give the full mathematical model in this paper but it is available alongside the open-source implementation.

\section{Experiments}\label{sec:expe}

We have implemented the ILP model-generation as an open-source tool called \texttt{jMCM}\footnote{Available on git: \url{https://github.com/remi-garcia/jMCM}}. We used the modeling language JuMP \cite{DunningHuchetteLubin_JuMPModelingLanguage_2017} to implement the model generator, which can then use any generic MILP solver supported by JuMP as backend. We chose the Gurobi 9.5.1 \cite{GurobiOptimizerReference_2020} solver executed with 4 threads on an Intel\textregistered{} Core\texttrademark{} i9-11950H CPU at 2.60GHz and a time limit of 30~minutes. We also fully automate a tool-chain with FloPoCo~\cite{DinechinPasca_DesigningCustomArithmetic_2011}, a state-of-the-art hardware code generator for arithmetic operators, to produce VHDL which we then synthesized for FPGA using Vivado v2018.2 for the xc7v585tffg1761-3 Kintex 7 device.

In this section, we will compare the impact of our models on different metrics, especially hardware metrics that are targeted only indirectly.
With these experiments, we want to confirm that minimizing the adder depth does positively impact the final hardware delay or that a focus on one-bit adders permits to reduce the hardware cost.
The research questions we answer in this section are summarized into four main questions:
\begin{itemize}[leftmargin=31pt]
	\item[(RQ1)] Does taking the adder depth into account lead to better hardware?
	\item[(RQ2)] Do adder graphs with less one-bit adders permit cost reduction in the final hardware?
	\item[(RQ3)] What is the impact of intermediate truncations?
	\item[(RQ4)] How increasing error bound can decrease the synthesized hardware costs?
\end{itemize}

To answer these questions, we solve all our models on benchmarks from image processing ($11$~instances) that have already been used to compare \MCM algorithms~\cite{Kumm_MultipleConstantMultiplication_2016_book,KummFanghaenelMoellerZipfMeyerBaese_FIRfilteroptimization_2013}.
We extend our tests on the whole FIRsuite project ($75$~instances)~\cite{FIRsuiteWeb}, which is a collection of digital filter designs and is a direct application of the MCM problem.
We solved multiple flavors of the \MCM problem, four to be precise, producing hundreds of adder graphs. Using FloPoCo and Vivado, we synthesized all these adder graphs and extracted the number of LUTs, the delay and power (we present the sum of the logic, dynamic and signal powers).
For conciseness, detailed results in \tablename~\ref{tab:allresults} are shown only for the image processing benchmark, while the statistics are discussed over the full set of $86$ benchmarks. The detailed report is available on git and results are reproducible in an automated way.

\subsection{Optimization settings and timings}

One of the advantages of ILP-based approach is that a ``good'' candidate solution is available after the first several second of resolution, even if optimality over a huge design space is too difficult to prove by a generic solver. And in most cases that we observed, the solver either proves the optimality within first 10 seconds, or timeouts. To be precise, for the metrics based on number of adders, we could prove optimality for two thirds of benchmarks, and for models targeting one-bit adders it decreases to only a third of proven optimal results. With the ILP modeling for MCM, the complexity of the problem is not necessarily in the number of target coefficients, as we could quickly solve large instances and struggle with smaller ones, but in the coefficients themselves. One limitation is, however, data representation in the ILP solvers and potential numerical instabilities that can arise. In our benchmarks, we detected that starting from $12$ bits, optimality could not be proven, but good candidate solutions were found for coefficients up to $19$ bits.

Since solving the \MCMA problem permits to quickly obtain adder graphs with a low number of adders, we use it as a warm-start for the \MCMB or \TMCM in order to speed up the resolution of complex models. And for any model, we give a bound on the number of adders by either using the RPAG library \cite{KummZipfFaustChang_Pipelinedaddergraph_2012}, if available, or the CSD greedy algorithm \cite{Bernstein_Multiplicationintegerconstants_1986}.

In many cases the MCM circuits are used in an iterative way, hence the output result is truncated to some intermediate format and then re-injected in the circuit, so that the input/output word lengths are the same.
For the image-processing benchmarks an 8-bit input/output word length is a reasonable choice, hence we limited the error to a single unit of least precision (ulp) of the output precision. When taking the error of last rounding into account, we obtain that the error bound for internal truncations is actually half an ulp.

\def\varheight{4.5cm}
\def\varshift{12pt}
\def\varenlarge{30pt}
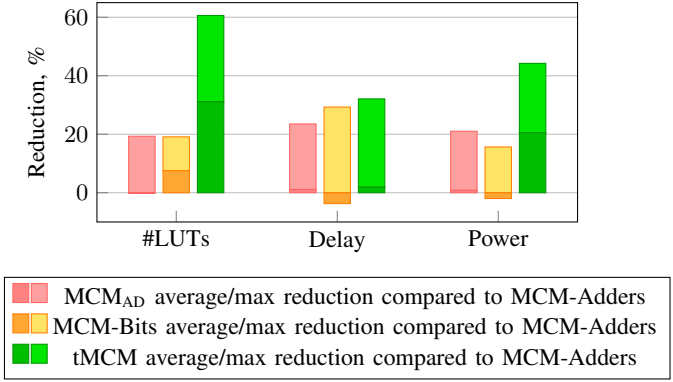
\begin{figure}
	\small
	\centering
	\begin{tikzpicture}
	\begin{axis}[
	ybar stacked,
	bar shift=0pt,
	ytick pos=left,
	xtick pos=bottom,
	legend style={at={(0.5,-0.25)}, /tikz/every even column/.append style={column sep=-2mm}, anchor=north, legend columns=2},
	enlarge x limits={abs=\varenlarge},
	ymin=-10,
	ymax=65,
	ytick={0,20,...,60},
	ymajorgrids=true,
	width=0.9\linewidth,
	height=\varheight,
	ylabel={Reduction, \%},
	symbolic x coords={\#LUTs, Delay, Power},
	xtick=data,
	point meta=explicit symbolic
	]
	\addlegendimage{mcmad, fill=mcmadfill}
	\addlegendentry{}
	\addlegendimage{mcmadmax, fill=mcmadfillmax}
	\addlegendentry{\MCMad average/max reduction compared to \MCMA}
	\addplot[legend entry=\phantom{.}, mcmb, fill=mcmbfill] coordinates {
		(\#LUTs, 7.54)(Delay, -3.73)(Power, -1.99)        };
	\addplot[legend entry=\MCMB average/max reduction compared to \MCMA, mcmbmax, fill=mcmbfillmax] coordinates {
		(\#LUTs, 11.560000000000002)(Delay, 29.29)(Power, 15.64)        };
	\addlegendimage{tmcm, fill=tmcmfill}
	\addlegendentry{}
	\addlegendimage{tmcmmax, fill=tmcmfillmax}
	\addlegendentry{\TMCM average/max reduction compared to \MCMA}
	\end{axis}
	\begin{axis}[
	ybar stacked,
	bar shift=-\varshift-2.5pt,
	hide axis,
	ytick pos=left,
	xtick pos=bottom,
	legend style={at={(0.5,-0.25)}, anchor=north, legend columns=1},
	every axis legend/.code={\let\addlegendentry\relax},
	enlarge x limits={abs=\varenlarge},
	ymin=-10,
	ymax=65,
	width=0.9\linewidth,
	height=\varheight,
	ylabel={Reduction, \%},
	symbolic x coords={\#LUTs, Delay, Power},
	xtick=data,
	point meta=explicit symbolic
	]
	\addplot[legend entry=\MCMad average reduction compared to \MCMA, mcmad, fill=mcmadfill] coordinates {
		(\#LUTs, -0.17)(Delay, 1.11)(Power, 0.82)        };
	\addplot[legend entry=\MCMad max reduction compared to \MCMA, mcmadmax, fill=mcmadfillmax] coordinates {
		(\#LUTs, 19.32)(Delay, 22.42)(Power, 20.2)        };
	\end{axis}
	\begin{axis}[
	ybar stacked,
	bar shift=\varshift,
	hide axis,
	ytick pos=left,
	xtick pos=bottom,
	legend style={at={(0.5,-0.25)}, anchor=north, legend columns=1},
	every axis legend/.code={\let\addlegendentry\relax},
	enlarge x limits={abs=\varenlarge},
	ymin=-10,
	ymax=65,
	width=0.9\linewidth,
	height=\varheight,
	ylabel={Reduction, \%},
	symbolic x coords={\#LUTs, Delay, Power},
	xtick=data,
	point meta=explicit symbolic
	]
	\addplot[legend entry=\TMCM average reduction compared to \MCMA, tmcm, fill=tmcmfill] coordinates {
		(\#LUTs, 31.11)(Delay, 1.86)(Power, 20.53)        };
	\addplot[legend entry=\TMCM max reduction compared to \MCMA, tmcmmax, fill=tmcmfillmax] coordinates {
		(\#LUTs, 29.5)(Delay, 30.230000000000004)(Power, 23.699999999999996)        };
	\end{axis}
	\end{tikzpicture}
	\caption{The cost reduction of proposed models compared to the corrected~\cite{Kumm_OptimalConstantMultiplication_2018}. Here \TMCM is for 8-bit inputs and faithful rounding to $50\%$ of the output word length.}\label{fig:averageandmaxgain}
\end{figure}

\subsection{RQ1: the impact of simultaneous adder count and adder depth minimization}

We start by analyzing the first improvement to the corrected \MCMA model, which is the simultaneous minimization of AD, as a secondary objective to the adder count. In $13$ out of $86$ benchmarks we exhibited a decrease of the AD by 1 and up to 4 stages, for the same minimal number of adders. Solving the \MCMA led to many adder graphs with AD greater than $4$ and up to $6$, while \MCMmAD shows that for every instance of our benchmarks there exists an adder graph with AD $\leq 4$. The running times were similar to our \MCMA, though some solutions were not proven optimal anymore.

Minimizing the AD is an empirical approach to lower the delay of resulting hardware, its area and power. In \figurename~\ref{fig:averageandmaxgain} we exhibit, with the red bars, the reductions in each metric compared to the corrected state-of-the-art \MCMA.
Over the $13$ instances impacted by the simultaneous AD minimization, the average improvements are small, around $1\%$ but can go up to roughly $20\%$ for all metrics.
We conclude that while \MCMA is already doing a good job w.r.t. the adder depth, it can be quite beneficial to automatically search the design space of adder graphs with minimal number of adders for the ones with least AD.

We also investigate the correlation between adder depth and the delay. We computed the correlation between different metrics and the hardware cost for all our benchmarks, each implemented in four flavors (\MCMA, \MCMmAD, \MCMB, \TMCM). As illustrated in \tablename~\ref{tab:correlation}, the correlation between the delay and the AD is $r=0.3481$. 
We find this quite interesting, since in previous works \cite{Kumm_OptimalConstantMultiplication_2018, Kumm_MultipleConstantMultiplication_2016_book} bounding the adder depth to some low value was motivated by supposedly lower delay, which is refuted by our analysis.
Indeed, as we observed, bounding the adder depth to a value lower than optimal, \ie{}, obtained with our \MCMmAD, always leads to a larger number of adders that have much bigger correlation with delay: $r=0.9549$.

As a result, we do not further investigate artificially bounding the AD but, by default, incorporate the AD minimization as a second objective in \MCMB and \TMCM.

\begin{table}[]
	\centering
	\caption{Correlation between different metrics and actual hardware cost}\label{tab:correlation}
	{\normalsize
		\begin{tabular}{@{}c|ccc@{}}
			& \#adders & \#one-bit adders & adder depth \\
			\noalign{\hrule height 0.5pt}
			\rule{0pt}{2.5ex}\#LUTs & $0.9602$ & $0.9967$ & $0.2367$ \\
			Delay & $0.9549$ & $0.913$ & $0.3481$ \\
			Power & $0.9812$ & $0.9659$ & $0.2592$ \\
		\end{tabular}
	}
\end{table}

\subsection{RQ2: hardware impact of the one-bit adder metric }

\begin{table*}[]
	\centering
	\caption{Detailed results for the image-processing benchmarks, where Delay is in (ns), Power is in (mW)}\label{tab:allresults}
	{\small
		\setlength{\tabcolsep}{4pt}
		\begin{tabular}{@{}ccccccc|cccccc|cccccc@{}}
			\toprule
			\multirow{2}{*}{Bench} & \multicolumn{6}{c}{\MCMA} & \multicolumn{6}{c}{\MCMB} & \multicolumn{6}{c}{\TMCM} \\
			\cmidrule(lr){2-7} \cmidrule(lr){8-13} \cmidrule(lr){14-19}
			& $N_A$ & $\AD$ & \#B & \#LUTs & Delay & Power & $N_A$ & $\AD$ & \#B & \#LUTs & Delay & Power & $N_A$ & $\AD$ & \#B & \#LUTs & Delay & Power \\ \midrule
			\multicolumn{1}{l|}{\texttt{G.3}} & 4 & 2 & 43 & 45 & 7.385 & 111 & 4 & 2 & 40 & 41 & 7.305 & 108 & 4 & 2 & 24 & 27 & 7.205 & 85 \\
			\multicolumn{1}{l|}{\texttt{G.5}} & 5 & 4 & 58 & 60 & 9.700 & 168 & 5 & 4 & 57 & 58 & 8.937 & 175 & 5 & 4 & 36 & 44 & 9.267 & 139 \\
			\multicolumn{1}{l|}{\texttt{HP5}} & 4 & 2 & 42 & 42 & 7.506 & 121 & 4 & 3 & 39 & 41 & 8.574 & 134 & 4 & 2 & 25 & 26 & 7.678 & 89 \\
			\multicolumn{1}{l|}{\texttt{HP9}} & 5 & 2 & 50 & 50 & 7.859 & 158 & 5 & 2 & 47 & 48 & 7.675 & 165 & 5 & 2 & 34 & 40 & 7.826 & 146 \\
			\multicolumn{1}{l|}{\texttt{HP15}} & 12 & 2 & 122 & 122 & 10.207 & 401 & 12 & 3 & 105 & 108 & 10.377 & 421 & 12 & 2 & 60 & 62 & 9.842 & 270 \\
			\multicolumn{1}{l|}{\texttt{L.3}} & 3 & 3 & 34 & 36 & 7.868 & 116 & 3 & 3 & 31 & 33 & 8.377 & 112 & 3 & 3 & 26 & 29 & 7.925 & 102 \\
			\multicolumn{1}{l|}{\texttt{LP5}} & 6 & 3 & 66 & 68 & 9.099 & 213 & 6 & 3 & 60 & 65 & 9.142 & 209 & 6 & 3 & 47 & 52 & 8.994 & 173 \\
			\multicolumn{1}{l|}{\texttt{LP9}} & 12 & 5 & 157 & 154 & 11.915 & 537 & 12 & 3 & 137 & 138 & 12.439 & 516 & 13 & 3 & 88 & 95 & 11.379 & 361 \\
			\multicolumn{1}{l|}{\texttt{LP15}} & 26 & 6 & 313 & 315 & 20.979 & 1359 & 27 & 4 & 250 & 258 & 15.617 & 1218 & 27 & 3 & 150 & 182 & 14.632 & 959 \\
			\multicolumn{1}{l|}{\texttt{U.3-1}} & 4 & 2 & 32 & 33 & 7.372 & 104 & 4 & 2 & 32 & 33 & 7.372 & 104 & 4 & 2 & 22 & 17 & 6.479 & 63 \\
			\multicolumn{1}{l|}{\texttt{U.3-2}} & 5 & 3 & 61 & 58 & 8.937 & 169 & 5 & 4 & 49 & 49 & 9.625 & 167 & 5 & 4 & 33 & 32 & 8.967 & 102 \\
			\bottomrule
		\end{tabular}
	}
\end{table*}

The detailed hardware results in \tablename~\ref{tab:allresults} demonstrate that minimizing the one-bit adders always reduces the number of LUTs, though in rare cases the results coincide with the \MCMA solution, \eg{}, for \texttt{U.3-1}. As illustrated by \figurename~\ref{fig:averageandmaxgain}, we reduced the number of LUTs by $\gainLUTsOBAvsNA\%$ on average, and $19.1\%$ at most for the full benchmark set.

As illustrated in \tablename~\ref{tab:allresults}, minimizing the one-bit adders can sometimes lead to a larger adder count, as in the case \texttt{LP15}, while significantly improving all hardware metrics. Also, in some cases, even if minimizing the AD is a default secondary objective, the optimal one-bit adder solutions require more stages.

To compare the impact of the one-bit adder metric vs. the adder count, we analyzed the correlation between each of them and the actual hardware cost. \tablename~\ref{tab:correlation} clearly demonstrates that the one-bit adders are most certainly in a linear relationship with the LUT count, with a correlation factor $r=0.9967$, which is stronger than $r=0.9602$ for the number of adders.
This does not mean that minimizing the number of one-bit adders will surely minimize the number of LUTs, yet it is reasonable to expect it.

The one-bit adder correlation factors for the delay and power are $0.9129$ and $0.9659$, respectively, which indicates a less strong relationship, which is also worse than for the number of adders. This analysis also confirms the detailed results in \tablename~\ref{tab:allresults}, where sporadic increase in delay and power for seemingly better \MCMB solutions can be observed. In general, the increase is small and could be neglected but is probably due to a different glitch path count, which can also be modeled and used to guide the adder topology~\cite{Kumm_OptimalConstantMultiplication_2018}.

With the above, we can conclude that using the one-bit adder metric is more beneficial for the LUT count, regardless of occasional increase in the adder depth and number of adders.

\subsection{RQ3 and RQ4: advantages of truncated adder graphs}

In~\cite{GarciaVolkovaKumm_TruncatedMultipleConstant_2022} we provide a preliminary analysis and comparison with the state-of-the-art models and our \TMCM by counting the number of one-bit-adders. On the same set of benchmarks, we deduced that the number of one-bit adders, was on average reduced by $\gainOBAsTvsOBA\%$ compared to \MCMB problem.

In this work we go further and synthesize all adder graphs for hardware. \tablename~\ref{tab:allresults} confirms that \TMCM provides major hardware cost reductions: for $8$-bit inputs/outputs, the number of LUTs is decreased by $\gainLUTsTvsOBA\%$, the delay by $\gaindelayTvsOBA\%$ and the power by $\gainpowerTvsOBA\%$.

These results are also expected due to the strong correlation factor between the LUT count and the number of one-bit adders, which are minimized in \TMCM. Similarly to the case of \MCMB, truncated adder graphs in some cases require more adders, which is natural, since there are many different corner cases that permit to reduce the one-bit adder count with a drawback of larger number of fundamentals. Interestingly, in presence of truncations the adder depth might be, on the contrary, smaller than for the \MCMB solutions, significantly improving power.

Of course, this significant performance improvement requires the embedded system designers to be able to provide an \emph{a priori} error bound for the outputs. However, we do not find it unreasonable, since analyzing the finite-precision behavior of the implemented system is expected for resource-constrained applications. Generic static analysis tools can be used to compute such bounds, such as~\cite{Daisy, Fluctuat}, or they can be manually deduced for specific applications, such as has been done for digital filters~\cite{VolkovaIstoanDeDinechinHilaire_TowardsHardwareIIR_2019}.

In the above discussion we fixed the accuracy of the MCM outputs by setting the output word length to $8$ bits, which is reasonable for an image-processing benchmark. In general, as the coefficients have different magnitudes, this meant that for some instances $8$-bit output contained the exact result, and for some only a fraction of it. In order to present a fair general comparison, we now vary the error bound for each instance and first remove a quarter, and then a half of the \textit{exact} output word length. For example, if the full-precision result requires $16$ bits, we solve \TMCM to faithfully round to $12$ bits and then to $8$ bits. In cases when the coefficient magnitude is close to that of the input, keeping half of the result bits is quite reasonable, as it basically corresponds to maintaining the same input/output size. This experiment also permits us to see how varying the error bound impacts the hardware cost and might help to find the sweet spot between accuracy and resources.

On average, reducing the output word length by $25\%$ results in $\gainOBAsTquartervsOBA\%$ one-bit adders reduction, compared to full precision. Leaving only $50\%$ of the output bits permits to significanly change the topology of the adder graphs (as illustrated by the adder depth and adder count change in) and obtain, on average, the reduction of $\gainOBAsThalfvsOBA\%$ in terms of one-bit adder count.
This significant improvement is also reflected in the actual hardware metrics, as illustrated in \figurename~\ref{fig:averageandmaxgain}, reaching in some cases a $61\%$ reduction in LUTs.

\section{Conclusion}

Many approaches exist for solving the \MCM problem but they mostly rely on solving the \MCMA problem, \ie{}, with a high-level metric.
In this work we propose a low-level metric based on counting the number of one-bit adders, and tackle the \MCM problem in different flavors: from minimization of the adder depth as a secondary objective, to adding intermediate truncations whilst computing a faithfully rounded output. With this work, the non-trivial \MCM design-space exploration for embedded system designers is automated and delegated to powerful ILP solvers, liberating the designers to study high-level questions, such as definition of the input/output word lengths for their application.

With the ILP-based approach, we showed that extending the \MCMA problem does not require too much effort, when the cost/constraint modeling is done, which opens an easy way for more extensions such as the glitch path count. Moreover, our MCM model can be incorporated into the design of more complex algorithms that use MCM, such as digital filters~\cite{FIRopt}.

We kept our focus on FPGA design but most of this work directly applies to ASICs.
As a perspective, refining the one-bit adder metric into half-adders and full-adders could further improve the results for ASICs.

Overall, we proposed a tool with options allowing for tackling the \MCM problem with respect to many end-user needs. Given enough time and numerical robustness of the MILP solver, the solving process permits to find optimal solutions and prove optimality with an exhaustive search, or propose candidate solutions if a timeout is fixed.

\bibliographystyle{IEEEtran}
\bibliography{mcm_submitted}

\end{document}